\documentclass[12pt]{article}
\usepackage{amsmath}
\usepackage{amstext}
\usepackage{amsfonts}
\usepackage{graphicx,epsfig}
\usepackage{indentfirst}
\begin{document}
\baselineskip 20pt
\title{Bell Inequalities in High Energy Physics\\[1.3cm]}

\begin{figure}
{\hspace{12.7cm}\vbox{\halign{&#\cr &GUCAS-CPS-07-03 \cr
&hep-ph/0702271\cr }}}
\end{figure}
\author{Yi-Bing Ding\footnote{ybding@gucas.ac.cn},\;
Junli Li\footnote{jlli04@mails.gucas.ac.cn},\; and
Cong-Feng Qiao\footnote{qiaocf@gucas.ac.cn}\\[0.5cm]
\small Dept. of Physics, Graduate
University, the Chinese Academy of Sciences  \\
\small {YuQuan Road 19A, 100049, Beijing, China}\\
}
\date{}

\maketitle

\begin{abstract}
We review in this paper the research status on testing the
completeness of Quantum mechanics in High Energy Physics, especially
on the Bell Inequalities. We briefly introduce the basic idea of
Einstein, Podolsky, and Rosen paradox and the results obtained in
photon experiments. In the tests of Bell inequalities in high energy
physics, the early attempts of using spin correlations in particle
decays and later on the mixing of neutral mesons used to form the
quasi-spin entangled states are covered. The related experimental
results in $K^0$ and $B^0$ systems are presented and discussed. We
introduce the new scheme, which is based on the non-maximally
entangled state and proposed to implement in $\phi$ factory, in
testing the Local Hidden Variable Theory. And, we also discuss the
possibility in generalizing it to the tau charm factory.
\end{abstract}

\noindent{\bf \hspace{1cm}PACS number(s):} 03.65.Fh, 14.40.Aq,
13.25.Gv \vspace{6pt}
\section{Introduction}
Quantum Mechanics (QM) is one of the most important foundations of
modern physics. However, the philosophic and physical debates on
this fundamental theory are still continuing ever since its first
presence. Among the various critiques on QM, the most important and
famous one is what proposed by Einstein and his collaborators on
whether the QM is a complete theory or not. Einstein, Podolsky, and
Rosen (EPR) \cite{epr} questioned the completeness of QM by using a
so-called \textit{Gedanken experiment} which was then named the EPR
paradox. In section 2 we introduce the EPR paradox in details, the
explanation for the paradox in local hidden variable theory (LHVT),
and the Bell theorem, which exhibits the contradiction of LHVT with
QM and presents the non-locality nature of QM as the foundation of
the modern quantum information theory. In section 3 we first
introduce some optical experiments in testing the Bell inequalities.
Then, we turn to the related studies in high energy physics in
section 4. The last section is remained for conclusions for the past
researches in testing the Bell Inequalities, and expectations for
future investigations, especially in high energy physics.

\section{From EPR to Bell inequalities}
\subsection{The EPR paradox}
In 1935, Einstein, Podolsky, and Rosen demonstrated in a work
\cite{epr} that quantum mechanics could not provide a complete
description for the ``physical reality" of two spatially separated
but quantum mechanically correlated particle system. In the paper
they described the following criterion of ``physical reality'':
\textit{if, without in any way disturbing a system, we can predict
with certainty (i.e., with probability equal to unity) the value
of a physical quantity, then there exists an element of physical
reality corresponding to this physical quantity.} Then they
proposed the necessary condition for theories to be complete:
\textit{every element of the physical reality must have a
counterpart in the physical theory.}

People noticed that the idealized experiment (\textit{Gedanken
experiment}) proposed by EPR is not suitable for designing the
practical experiment. It requires to configure an entangled state,
which is of the eigenstate of relative position and total momentum.
However, this is not practical. Moreover, even if it could be
constructed, such state cannot be a stationary state. It will only
be in transitory existence, which makes the EPR argument fail.

Bohm \cite{bohm} proposed a more realistic experiment which can
illustrate the EPR paradox. He considered the two-particle
spin-one-half system in spin singlet and zero angular momentum. in
spin space, the wave function of this state can be expressed as
\begin{eqnarray}
|\Psi\rangle = \frac{1}{\sqrt{2}} (|+\rangle_A |-\rangle_B -
|-\rangle_A |+\rangle_B )\; ,
\end{eqnarray}
where the single particle states $|+\rangle$ and $|-\rangle$ denote
``spin up" and ``spin down" in certain coordinate frame. Assuming
the two particle interaction does not involve spin-dependent term,
particles are allowed to separate apart with the total spin of the
system invariant, for example along the $y$ direction. When they are
separated well beyond the range of interaction, we can measure the
$z$-component of the spin of particle A. Due to angular moment
conservation at all time, we can predict that the $z$-component of
spin B must have the opposite value. In the meantime, because the
spin singlet has spacial rotation invariance, the same thing happens
when we measure the $x$-component of spin of particle A. Since the
two particles are far apart with each other, the locality condition
guarantees that the particle B does not know what happens to A while
the measurement performs. Therefore, it shows that the B particle
spins along $x$ and $z$ axes should be both physical realities. In
QM the spin operators along different axes do not commute and thus
can not simultaneously have definite values, therefore they can not
be simultaneously in physical realities. Hence, Einstein concluded
that QM must be incomplete.

Bohr contested not the EPR demonstration but their premises. His
point of view is that an element of reality is associated with a
concretely performed act of measurement. We can not perform the
measurement along different axes simultaneously on particle A, so
the spins of the particle B along different axes need not to be
simultaneously physical realities. However, as Einstein questioned
that these arguments make the reality of particle B depend upon the
process of the measurement performed on the first particle, and he
believed that \textit{``no reasonable definition of reality could be
expected to do this."}

\subsection{Bell inequalities}

To avoid the EPR paradox, it might be a reasonable choice to
postulate some additional `hidden variables', which presumably will
restore the completeness, determinism and causality to the theory.
This kind of theories are named the local hidden variable theories.
Nevertheless, once von Neumann, based on some axioms \cite{neumann},
demonstrated that it is impossible to construct such a hidden
variable theory \cite{nonlocality} reproducing all the results of
QM. It was later on discovered that one of the von Neumann's axioms
in getting his conclusion is too much restrictive. And, indeed some
counter examples were constructed in the two dimensional space
\cite{revbell}. That means the LHVT model can produce all the QM
predictions but without fulfil von Neumann's restrictive hypotheses.
Nevertheless, there remains certain difference in between QM and
LHVT. In 1964 Bell showed \cite{bell} that in realistic LHVTs the
two particle correlation functions satisfy a set of Bell
inequalities (BI), whereas the corresponding QM predictions may
violate these inequalities in some region of parameter space. The
definition of correlation for LHVTs and QM, according to Bohm, read
respectively as:
\begin{eqnarray}
E(\textbf{a},\textbf{b}) & = & \int \mathrm{d}\lambda
\rho(\lambda)
A(\textbf{a},\lambda)B(\textbf{b},\lambda)\; , \\
E(\textbf{a},\textbf{b}) & = & \langle \psi| \sigma \cdot
\textbf{a} \otimes \sigma \cdot \textbf{b} |\psi \rangle =
-\textbf{a} \cdot \textbf{b}\; . \label{bohm}
\end{eqnarray}
Here, $\rho(\lambda)$ is the distribution of hidden variable
regardless of whether $\lambda$ is a single variable or a set, or
even a set of functions. These variables can be either discrete or
continuous. $\textbf{a}$ and $\textbf{b}$ indicate spin
directions. The original inequality obtained by Bell is
\begin{eqnarray}
|E(\textbf{a},\textbf{b}) - E(\textbf{a},\textbf{c})| -
E(\textbf{b},\textbf{c}) \leq 1\; , \label{bell1}
\end{eqnarray}
where $\textbf{a}, \textbf{b}, \textbf{c}$ mean three different
spin directions. In 1969, Clauser, Horne, Shimony, and Holt (CHSH)
\cite{CHSH} generalized the inequality (\ref{bell1}) to a more
practical one, i.e.
\begin{eqnarray}
S=|E(\textbf{a},\textbf{b}) - E(\textbf{a},\textbf{b}')| +
E(\textbf{a}',\textbf{b}) + E(\textbf{a}',\textbf{b}')
 \leq 2\; .
\end{eqnarray}
A similar inequality to CHSH was derived by Bell in 1971
\cite{bell1971}, read as
\begin{eqnarray}
S = |E(\textbf{a},\textbf{b}) - E(\textbf{a},\textbf{b}')| +
|E(\textbf{a}',\textbf{b}) + E(\textbf{a}',\textbf{b}')|
 \leq 2\; . \label{bell}
\end{eqnarray}
The correlation function $E$ in above inequalities is defined as
\begin{eqnarray}
E(\textbf{a},\textbf{b}) = P_{++}(\textbf{a},\textbf{b}) -
P_{+-}(\textbf{a},\textbf{b}) - P_{-+}(\textbf{a},\textbf{b}) +
P_{--}(\textbf{a},\textbf{b})\; ,
\end{eqnarray}
where $P_{\pm\pm} = N_{\pm\pm}(\textbf{a},\textbf{b})/N$, $N$ is the
total number of particle pairs, and $N_{++(+-)}$ means that two
particle has the same (opposite) spin directions. To suffice for
experimental test, the total number of particle pair emissions $N$
should be known. However, in real practice the probability cannot be
measured without either destroying or depolarizing the particle
pairs. In 1974 Clauser and Horne (CH) \cite{CH} deduced an
inequality, for which the upper limit is experimentally testable
without knowing the $N$. That is
\begin{eqnarray}
-1 \leq P_{++}(\textbf{a},\textbf{b}) -
P_{++}(\textbf{a},\textbf{b}') + P_{++}(\textbf{a}',\textbf{b}) +
P_{++}(\textbf{a}',\textbf{b}') - P_{++}(\textbf{a}',\infty) -
P_{++}(\infty,\textbf{b}) \leq 0, \label{CH}
\end{eqnarray}
where $P_{++}(\infty,\textbf{b})$ denotes the probability of finding
a pair of particles with no polarization detection on one side. It
is easy to find that the CH inequality (\ref{CH}) is consistent with
inequality (\ref{bell}). Provided that in an experiment with two
detectors and double channel analyzers, one can get three similar
sets of inequalities like Eq.(\ref{CH}) with different indices
$P_{-+}, P_{+-}, P_{--}$. Multiplying the inequalities with $P_{-+}$
and $P_{+-}$ by $-1$ and combining these four inequalities we can
obtain the inequality (\ref{bell}).

It is generally realized that unlike the von Neumann's mathematical
results these inequalities can be reached in experiment in testing
the validity of QM in comparison with LHVTs.

\subsection{Generalizations of Bell theorem}

Bell theorem reveals peculiar properties of quantum ``entangled"
states that were previously not appreciated. Many a generalization
of Bell inequality aiming at getting optimal violations was
developed. Better inequalities (inequalities with larger violation
and/or wide range of parameter space for violation) are of both
experimental and theoretical interest. In further development, one
may create new inequalities, or explore the non-local character of a
particular quantum state. Of course these two seemingly different
investigation schemes are correlated.

Braunstein and Caves \cite{chain} made an extension of the
inequality (\ref{bell}). They added two kinds of (\ref{bell}) up
with different directions and got:
\begin{eqnarray}
S = |E(\textbf{a},\textbf{b}'') + E(\textbf{b}'',\textbf{a}'') +
E(\textbf{a}'',\textbf{b}') + E(\textbf{b}',\textbf{a}')
+E(\textbf{a}',\textbf{b}) - E(\textbf{b},\textbf{a})| \leq 4\; .
\label{bcaves}
\end{eqnarray}
Usually, combining two inequalities directly will lead to an
inequality with looser constraint than before. However Ref.
\cite{chain} demonstrated that this kind of adding chain may lead to
even stronger quantum violations. In this way, we may reexpress
their result in a different form:
\begin{eqnarray}
S_N = N |E(\pi/N)| \leq N-2\; ,\label{limitation}
\end{eqnarray}
where $N \geq 3$. It is very interesting to notice that when $N =
3$ (\ref{limitation}) corresponds to the maximal violation of
(\ref{bell1}); $N = 4$ corresponds to the maximal violation of
(\ref{bell}); and $N = 6$ corresponds to the maximal violation of
(\ref{bcaves}). Taking $\pi/N = \theta$, we have
\begin{eqnarray}
|E(\theta)| \leq 1 - \frac{2\theta}{\pi}\; ,
\end{eqnarray}
which is similar to the Eq.(2.5) of \cite{tqvistbook}. Braunstein
and Caves also put forward the idea of information-theoretic Bell
inequalities \cite{inforbell1}. The information-theoretic Bell
inequalities was derived from the classical Shannon entropy and are
violated by the quantum mechanical EPR pairs. This makes it possible
to use the information theory to study the separability and
nonlocality of quantum states. For more details, readers should
refer to references
\cite{inforbell2,inforbell3,inforbell4,inforbell5}.

In 1989, Greenberger, Horne, and Zeilinger (GHZ) \cite{ghz2,ghz3}
showed that for certain three and four particle entangled states
there exists a conflict of QM prediction with local realism even for
perfect correlation. That is the LHVT and QM can both make definite
but opposite predictions. In 1992 Hardy proved \cite{hardy}, without
using inequalities, this kind of definite confliction can occur for
any non-maximally entangled state composed of two two-level
subsystems. Later on Hardy's argument was improved by Jordan
\cite{jordan}. He demonstrated that there exist four projection
operators satisfying
\begin{eqnarray}
\langle FG \rangle = 0\; , & & \langle D(1-G) \rangle = 0\;
,\label{jordan0}\\ \langle (1-F)E \rangle = 0\; , & & \  \ \ \langle
DE \rangle
> 0\; , \label{jordan}
\end{eqnarray}
which are in contradiction with LHVTs. In above and following
equations, the alphabetic letter on the left side represents the
projector of particle 1, and the right one for particle 2.
Eq.(\ref{jordan0}) and Eq.(\ref{jordan}) can be easily understood,
i.e., if $D=1$ then $G=1$ according to the second equality of
Eq.(\ref{jordan0}). And similarly if $E=1$ then $F=1$ according to
the first equality of Eq.(\ref{jordan}). From the second inequality
of Eq.(\ref{jordan}) we can infer that it is possible for $D$ and
$E$ to be 1 simultaneously, and so are the $F$ and $G$. However,
this is apparently in confliction with what the first equality of
Eq.(\ref{jordan0}) tells. Jordan also demonstrated in a converse way
\cite{jordan} that for any choice of four different measurements,
there exists a state satisfying Hardy's argument. Garuccio in 1995
found \cite{garuccio} that the contradiction between QM and LHVT can
be embedded in CH inequalities of (\ref{CH}), i.e.
\begin{eqnarray}
\langle DE \rangle \leq \langle FG \rangle + \langle D(1-G)
\rangle + \langle (1-F)E \rangle\; . \label{eber}
\end{eqnarray}
Along Hardy's logic, Cabello \cite{cabello0} formulated a GHZ type
of proof involving just two observers. Ref.\cite{cabello1}
demonstrated that for the state that is a product of two singlet
states, there exists a operator satisfying $F_{QM} = \langle \Psi| O
|\Psi\rangle = 9$ and $F_{LHVT} \leq 7$, which is obviously
inconsistent. For recent developments on this respect one can find
in the series works of Cabello's \cite{cabello2,cabello3,cabello4}.

Actually, the investigations on non-locality and the violation of
Bell inequalities are not so transparent as explained above,
especially when the mixed states and multi-particle high dimension
systems are concerned. Since, it is not our main focus of this
article, we suggest interested readers to refer to a recent review
\cite{Genovese} and references therein.

\section{Bell inequalities in optical experiment}

Many experiments in regard of the Bell inequalities have been
carried out by using the entangled photons. In the optical
experiment the correlation of polarizers in orientations
\textbf{a} and \textbf{b} is defined as follows:
\begin{eqnarray}
E(\textbf{a}, \textbf{b}) = \frac{N_{++}(\textbf{a}, \textbf{b}) +
N_{--}(\textbf{a}, \textbf{b}) - N_{+-}(\textbf{a}, \textbf{b}) -
N_{-+}(\textbf{a}, \textbf{b}) }{N_{++}(\textbf{a}, \textbf{b}) +
N_{+-}(\textbf{a}, \textbf{b}) + N_{-+}(\textbf{a}, \textbf{b}) +
N_{--}(\textbf{a}, \textbf{b})}\; ,
\end{eqnarray}
where $N_{+-}$ is the coincidence rate of photon polarizations; $+$
for parallel and $-$ for perpendicular to the chosen direction. Of
the various optical experiments, one of the important ones was
carried out by Aspect \textit{et al.} \cite{aspect1}, in which the
photons are generated from the atomic cascade radiation $J = 0
\rightarrow J = 1 \rightarrow J = 0$. In the experiment they use the
two-channel polarizers in orientations \textbf{a} and \textbf{b},
and a fourfold coincidence counting system by which the four
coincidence rates $N_{\pm\pm}(\textbf{a}, \textbf{b})$ can be
measured in a single run, and they obtain directly the polarization
correlation $E(\textbf{a}, \textbf{b})$. Their measurement gave
\begin{eqnarray}
S_{exp} = 2.697\pm 0.015\; .
\end{eqnarray}
This result is in excellent agreement with the predictions of
quantum mechanics, which, for their polarizer efficiencies and lens
apertures, gives $S_{QM} = 2.7\pm 0.05$. This experiment has been
performed with the static setups in which polarizers are fixed for
the whole duration of a run. A more important improvement of this
experiment was made by the same group of people \cite{aspect2}, in
which they added two optical switches that can be randomly chosen in
between two directions. The result also violates the upper limit of
Bell's inequality and in a good agreement with QM calculation.

Some other relevant and important progresses in this direction were
realized by using the parametric down-conversion (PDC)
\cite{pdc1,pdc2} technique in generating the entangled photon pairs.
An ideal experiment with two channel polarizers, which randomly
reoriented during the propagation of photons, has been fulfilled in
real world \cite{weihs}. The necessary space-like separation of the
observation was achieved by keeping sufficiently large physical
distance between the measurement stations (Alice and Bob was
spatially 400m apart in the experiment), by ultra-fast and random
setting of the analyzers, and by completely independent data
registration. The experiment finally gave
\begin{eqnarray}
S_{exp} = 2.73 \pm 0.02
\end{eqnarray}
for 14700 coincident events collected in 10s. This correspond to
violation of the CHSH inequality of 30 standard deviations assuming
only the statistical errors exist.

Recent measurements of the Bell inequality violation are realized
through the multi-photon entangled states
\cite{multiphoton1,multiphoton2}. It is directly applied to test the
multi-photon generalizations of the Bell theorem
\cite{multiphoton3}. The experimental result complies with the
quantum mechanics prediction while contradicts with the LHVTs
prediction by over 8 standard deviations \cite{multiphoton3}.

The non-maximally entangled Hardy state was also realized in optical
experiment \cite{Hardyphoton}. The measurement further confirmed the
QM but denied local realistic results \cite{Hardyphoton}. A
generalization \cite{cabellogen} of Cabello's argument in
Ref.\cite{cabello1} was put into experiment in \cite{cabelloexp},
using two photon four dimension entanglement (two polarization and
two spatial degrees of freedom). The observable $F_{QM} = \langle
\Psi| O |\Psi\rangle = 8.56904 \pm 0.00533$ gave a violation of
LHVTs by about 294 standard deviations.

In all, for now all of the known experimental results
\cite{aspect1,aspect2,weihs,multiphoton3,Hardyphoton,cabelloexp} in
photon experiment are substantially consistent with the prediction
of the standard QM. It is well-known that the main difficulty in the
photon experiment is of the detect efficiency. Although the
situation is improved in the PDC case, in practice the efficiency is
still quite low. For example, the detection/collection efficiency is
only abut 5\% in \cite{weihs}. As aforementioned the total number of
emission is very important to the setup of correlation. To make
these experimental measurements logically comparable to Bell
inequalities one needs to make supplementary assumptions. That is:
the ensemble of actually detected pairs is independent of the
orientations of the polarimeters, and the detected photon pairs is a
fair sample of the the ensemble of all emitted pairs. In the
multi-photon case the similar detection loophole appears as well
\cite{Genovese}.

\section{Bell inequalities in High Energy Physics}
\subsection{Motivations and some early attempts}
People notice that the former experiments in testing the
completeness of QM are mainly limited in the electromagnetic
interaction regime, i.e., by employing the entangled photons, no
matter whether the photons are generated from atomic cascade or PDC
method. Considering of the fundamental importance of the concerned
question, to test the LHVT in experiment with massive quanta and
with other kinds of interactions is necessary \cite{abel}.

To this aim, the spin singlet state, as first advocated by Bohm
and Aharonov \cite{bohm} to clarify the EPR argument, is exploited
in experiment at the beginning. Lamehi-Rachti and Mitting
\cite{wmitting} performed an experiment in the low energy
proton-proton scattering at Saclay tandem accelerator. Their
measurement of the spin correlation of protons gave a good
agreement with what the QM tells.

As early as 1960s the EPR-like features of the $K^0\bar{K}^0$ pair
in the decays of $J^{PC}=1^{--}$ vector particles were noticed by
some authors \cite{leeyang,inglis,day,lipkin}. In the early attempts
of testing LHVTs through the Bell inequality in high energy physics,
people focused on exploiting the nature of particle spin
correlations \cite{abel,tqvist,privitera}. Typically, in Ref.
\cite{tqvist} T\"ornqvist suggested to measure the BI through the
following process:
\begin{eqnarray}
e^+e^-\rightarrow\Lambda\bar{\Lambda}\rightarrow\pi^-\; p\;
\pi^+\; \bar{p}\; .
\end{eqnarray}
Two different decay modes of $\eta_c \rightarrow \Lambda
\bar{\Lambda}$ and $J/\psi \rightarrow \Lambda \bar{\Lambda}$ are
considered by T\"orquist. In Ref.\cite{tqvistbook} the matrix
element for $\eta_c$ or $J/\psi$ decay generically takes the
following form:
\begin{eqnarray}
A \propto \sum_{ij} \langle \chi_p| \textbf{M}_a| \chi_{\Lambda_i}
\rangle s_{i\!j} \langle \chi^{\dagger}_{\bar{\Lambda}_j}|
\textbf{M}^{\dagger}_b |\chi^{\dagger}_{\bar{p}} \rangle\; .
\end{eqnarray}
Here, $M_{i}$ represents the interaction which induces the hadronic
transition of $\Lambda$ to final states. $s_{ij}$ represents spin
structure of the charmonium. After taking the standard procedure,
one obtain the transition probability. For example, for $\eta_c$
decay it reads:
\begin{eqnarray}
R(\hat{\textbf{a}},\hat{\textbf{b}}) \propto 1+\alpha^2
\hat{\textbf{a}}\cdot\hat{\textbf{b}}\; , \label{ratioeta}
\end{eqnarray}
where $\alpha$ denotes the $\Lambda$ decay asymmetry parameter;
$\hat{\textbf{a}}$ and $\hat{\textbf{b}}$ are unit vectors along the
$\pi^+$ and $\pi^-$ momenta in $\bar{\Lambda}$ and $\Lambda$ rest
frame, respectively. T\"ornqvist argued that apart from the constant
$\alpha^2$ and the sign before
$\hat{\textbf{a}}\cdot\hat{\textbf{b}}$, (\ref{ratioeta}) is in
equivalence with Eq.(\ref{bohm}) obtained in measuring the spin
correlation in the Bohm's \textit{Gedanken experiment}. Here, the
directions of the pion momentum $\hat{\textbf{a}}$ and
$\hat{\textbf{b}}$ take the place of the spin-analyzing directions
of the polarimeters.

For $J/\psi$ decays,
\begin{eqnarray}
R(\hat{\textbf{a}},\hat{\textbf{b}}) \propto 2 ( 1 -
\frac{k^2}{E^2_{\Lambda}} \sin^2\theta )(1 - \alpha^2
\hat{a}_n\hat{b}_n) + \frac{k^2}{E^2_{\Lambda}} \sin^2\theta [1 -
\alpha^2(\hat{\textbf{a}}\cdot\hat{\textbf{b}} -
2\hat{a}_x\hat{b}_x)]\; .\label{jpsi}
\end{eqnarray}
The DM2 Collaboration \cite{dm2} observed $7.7 \times 10^6\ J/\psi$
events with about $10^3$ being identified as from process $J/\psi
\rightarrow \Lambda \bar{\Lambda} \rightarrow \pi^-p\pi^+\bar{p}$.
The experimental measurement unfortunately does not give a very
significant result\cite{tqvistbook} due to the insufficient
statistics.

A similar process was suggested by Privitera \cite{privitera},
i.e.,
\begin{eqnarray}
e^+e^- \rightarrow \tau^+ \tau^- \rightarrow \pi^+
\bar{\nu}_{\tau} \pi^-\nu_{\tau}\; .
\end{eqnarray}
In analogy with what in charmonium decays, in this case the
expected correlation rate is given by
\begin{eqnarray}
N(\hat{\textbf{p}_1},\hat{\textbf{p}_2}) \propto
1-\frac{1}{3}\hat{\textbf{p}_1} \cdot \hat{\textbf{p}_2}\; ,
\label{tau}
\end{eqnarray}
where $\hat{\textbf{p}}_1$ and $\hat{\textbf{p}}_2$ are unit
vectors in the momentum directions of $\pi^+$ and $\pi^-$,
respectively. Hereby, the strong spin correlation between two
$\tau$'s reveals the nonlocal nature of the EPR argument. The
subsequent $\tau$ decay works as a spin analyzer, and the
correlation is transferred to the decay products.

The above mentioned designs for experimentally measuring the
violation of BI are delicate and attractive, however, people found
that such proposals possess controversial assumptions \cite{book}.
They all assume that the decay matrix elements contain the
nonlocal correlations, i.e., Eq.(\ref{ratioeta}, \ref{jpsi},
\ref{tau}). However, there is no dichotomic observable which can
be directly measured in real experiment. The momentum of pion is a
continuous variable, and different momenta are compatible, i.e.
[$(\hat{P}_{\pi^+})_i$, $(\hat{P}_{\pi^-})_j$] = 0
\cite{abel,book}. Thus a LHVT can be constructed in respect of all
the results from QM, and there will be no violation of Bell
inequality anyway.

\subsection{Testing correlation by virtue of quasi-spin}

In testing the LHVTs in high energy physics, using the ``quasi-spin"
to mimic the photon polarization in the construction of entangled
states is a practical way. For example, for kaon the quantum number
of strangeness $S$, which takes either 1 or -1, can play the role of
spin. Several groups suggested to study the $K^0\bar{K}^0$ system in
the $\phi$ factory to test the LHVT (for details, see Ref.
\cite{Bertlmann} and references therein). Up to now, there are two
different ways in the ``quasi-spin" scheme. In the first way, one
fixes up the quasi-spin, but leaves the freedom in time. For
example, one measures the Flavor Taste in different decaying time on
each side, then the time differences plays the role of polarization
angles. The second one is to leave the freedom in quasi-spin but to
fix the time. In this case we measure the different eigenstates of
the particles at the same time on each side, then the different
eigenstates play the role of polarization angles.

A typical process produces entangled state in $K^0 \bar{K}^0$ system
is through $ e^+e^- \rightarrow \phi \rightarrow K^0 \bar{K}^0$. The
wave function of the $J^{PC} = 1^{--}$ particles, like $\phi$ which
decays into $K^0 \bar{K}^0$, can be formally configured as
\cite{cptviolation}:
\begin{eqnarray}
|\phi\rangle = \frac{1}{\sqrt{2}}\{ |K^0\rangle |\bar{K}^0\rangle
- |\bar{K}^0\rangle |K^0\rangle \}\; . \label{kaonentangle}
\end{eqnarray}
Similar expressiones apply to $\Upsilon(4S) \rightarrow B^0
\bar{B}^0$, $\Upsilon(5S) \rightarrow B_{s}^0 \bar{B}_{s}^0$, and
$\psi(3770)\rightarrow D^0 \bar{D}^0$ cases.

In the following we explain the above mentioned techniques in a
bit details. First, we consider the situation in which the meson
state takes place of spin polarization discussed in preceding
sections. For kaon system, there are three different kinds of
eigenstates, those are: the mass, $CP$, and Strangeness.

We define the effect of $\hat{C}\hat{P}$ operators acting on the
$K^0$ and $\bar{K}^0$ states, like
\begin{eqnarray}
\hat{C}\hat{P} |K^0 \rangle &=& |\bar{K}^0 \rangle\; ,
\\ \hat{C}\hat{P} |\bar{K}^0 \rangle &=& |K^0 \rangle\; ,
\end{eqnarray}
up to an arbitrary phase. With this choice in phase the $CP$
eigenstates can be expressed as:
\begin{eqnarray}
|K^0_1\rangle & = & \frac{1}{\sqrt{2}}\{ |K^0\rangle +
|\bar{K}^0\rangle \}\; , \\ |K^0_2\rangle & = &
\frac{1}{\sqrt{2}}\{ |K^0\rangle - |\bar{K}^0\rangle \}\;
.\label{kaon1}
\end{eqnarray}
And correspondingly the mass eigenstates are:
\begin{eqnarray}
|K_S\rangle & = & \frac{1}{N}\{p|K^0\rangle +
q|\bar{K}^0\rangle\}\; , \\|K_L\rangle & = &
\frac{1}{N}\{p|K^0\rangle - q|\bar{K}^0\rangle\}\; ,\label{kaon2}
\end{eqnarray}
where $p=1+\varepsilon,q=1-\varepsilon$, and $N^2 = |p|^2 +
|q|^2$. The $\varepsilon$ is the normal $CP$ violation parameter.
With above knowledge, Eq.(\ref{kaonentangle}) can be reexpressed
as:
\begin{eqnarray}
|\phi\rangle & = & \frac{1}{\sqrt{2}}\{ |K_2\rangle |K_1\rangle -
|K_1\rangle |K_2\rangle \}\; ,\\ |\phi\rangle & = &
\frac{N^2}{2\sqrt{2}pq}\{ |K_L\rangle |K_S\rangle - |K_S\rangle
|K_L\rangle \}\; . \label{kaon-no-cp}
\end{eqnarray}
It is more convenient to use Wigner's inequality which can be
derived from Eq. (\ref{bell1}) \cite{Bertlmann}, i.e.:
\begin{eqnarray}
P(\textbf{a}, \textbf{b}) \leq P(\textbf{a}, \textbf{c}) +
P(\textbf{b}, \textbf{c})\; , \label{cpviolation}
\end{eqnarray}
where $\textbf{a},\textbf{b}, \textbf{c}$ are the same as in
Eq.(\ref{bell1}) and $P$s represent the probabilities with
subscripts in (\ref{CH}) suppressed. According to Ref.
\cite{Uchiyama}, we chose the following states as the quasi-spin:
\begin{eqnarray}
\textbf{a} & = & |K_S\rangle \; ,\\ \textbf{b} & = & |K^0\rangle
\; ,\\ \textbf{c} & = & |K_1\rangle \; .
\end{eqnarray}
Then, the inequality (\ref{cpviolation}) turns to be
\begin{eqnarray}
P(m_S, S=+1) \leq P(m_S, CP+) + P(CP+,S=+1)\; ,
\label{wigner_kaon}
\end{eqnarray}
where $P(m_S,S=+1)$ means the coincident rate while on one side it
is found to be $K_S$ and on the other side is found to be $K^0$.
The same notation applies to $P(m_S, CP+)$ and $P(CP+,S=+1)$.
Substitute (\ref{kaon1}) and (\ref{kaon2}) into
(\ref{wigner_kaon}), the inequality becomes \cite{Uchiyama}:
\begin{eqnarray}
\rm{Re}\{ \varepsilon\} \leq |\,\varepsilon|^{2}, \label{cpin}
\end{eqnarray}
which is obviously violated by the experimental measurements on
$\varepsilon$ \cite{PDG}. It is interesting to notice that as
$\textbf{b}$ taken to be $|\bar{K}^0\rangle$, (\ref{cpviolation})
becomes $-\rm{Re}\{ \varepsilon\} \leq |\,\varepsilon|^2$
\cite{Hiesmayr}, and it will be always true. As pointed out in
Ref.\cite{Bertlmann}, the (\ref{cpin}) is taken at the beginning
time, when the entangled kaon pairs are not well separated; what
tested is only the contextuality rather than non-locality.

As mentioned in above, we can also choose different time to measure
the final states, the kaons, on each side. For illustration, we
choose the quantum number of Strangeness as the quasi-spin in our
consideration, but neglect the CP violation effect, which in some
sense is a good approximation.

With the time evolution, the initial entangled state, like in
(\ref{kaonentangle}), becomes:
\begin{eqnarray}
|\Psi(t_l, t_r)\rangle = \frac{1}{\sqrt{2}} \{ e^{-i(m_Lt_l +
m_St_r)}e^{-\frac{\Gamma_L}{2}t_l-\frac{\Gamma_S}{2}t_r}|K_L\rangle
|K_S\rangle \nonumber \\ - e^{-i(m_St_l +
m_Lt_r)}e^{-\frac{\Gamma_S}{2}t_l-\frac{\Gamma_L}{2}t_r}|K_S\rangle
|K_L\rangle\}\; .
\end{eqnarray}
Here in above, the small letters $l$ and $r$ denote left side and
right side, suppose we name the two entangled particles to be left
and right without lose generality. Chooseing different measurement
time for two sides, we have the coincident rate
\cite{quant-ph/0501069}:
\begin{eqnarray}
P(K^0,t_l;K^0,t_r) & = & P(\bar{K}^0,t_l;\bar{K}^0,t_r) \nonumber
\\ & = & \frac{1}{8}\{ e^{-\Gamma_Lt_l-\Gamma_St_r} + e^{-\Gamma_St_l-\Gamma_Lt_r}
 - 2e^{-\frac{\Gamma_L+\Gamma_S}{2}(t_l+t_r)} \cos(\Delta m\Delta t)
 \}\; , \\ P(K^0,t_l;\bar{K}^0,t_r) & = & P(\bar{K}^0,t_l;K^0,t_r) \nonumber
\\ & = & \frac{1}{8}\{ e^{-\Gamma_Lt_l-\Gamma_St_r} + e^{-\Gamma_St_l-\Gamma_Lt_r}
 + 2e^{-\frac{\Gamma_L+\Gamma_S}{2}(t_l+t_r)} \cos(\Delta m\Delta t)
 \}\; .
\end{eqnarray}
Here, $P(K^0(\bar{K}^0),t_l;K^0(\bar{K}^0),t_r)$ represents the
probability of finding $K^0(\bar{K}^0)$ on the left side at time
$t_l$ and $K^0(\bar{K}^0)$ on the right side at time $t_r$. The
expectation value of correlation is:
\begin{eqnarray}
E(t_l,t_r) = - \cos(\Delta m\Delta t)
e^{-\frac{\Gamma_L+\Gamma_S}{2}(t_l+t_s)}\; .
\label{kaoncorrelation}
\end{eqnarray}
Inserting this correlation directly to CHSH inequality, one can
immediately find that the violation of inequality depend on the
ratio of $x = \Delta m/\Gamma$ \cite{Bertlmann}, where the $\Delta
m$ characterizes of Strangeness oscillation and the $\Gamma$
characterizes the weak decays. For the case of $x$ being small, that
means the oscillation is dominated by the weak decays, there will be
no violation of CHSH inequalities. Among the known neutral mesons,
only $B^0_S \bar{B}^0_S$ system has a big enough experimental value
of $x$, and hence the violation of inequalities might be found there
\cite{Bertlmann}.

The EPR-type Strangeness correlation in the process
$p\bar{p}\rightarrow K^0\bar{K}^0$ has been tested at the CPLEAR
detector \cite{cplear} at CERN. In the experiment the
$K^0\bar{K}^0$ pairs were created in $J^{PC} = 1^{--}$
configuration. The wave function at proper time $t_l = t_r = 0$ is
\begin{eqnarray}
|\Psi(0, 0)\rangle = \frac{1}{\sqrt{2}} [|K^0\rangle_l
|\bar{K}^0\rangle_r - |\bar{K}^0\rangle_l  |K^0\rangle_r]\; .
\end{eqnarray}
\begin{figure}[t,m,u]
\begin{center}
\includegraphics[width=8cm,height=5cm]{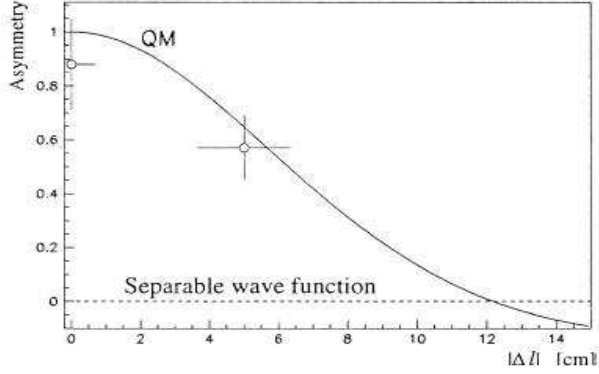}
\end{center}
\caption{\label{cplear} \small The best fit to the experimental
measurement \cite{cplear}. The two points with error bars
correspond to time difference $\Delta t = 0$ and $\Delta t = 1.2
\tau_s$. The solid line represents the QM prediction.}
\end{figure}
In the experiment, two kinds of measurements were performed. The
first one was to perform the measurements on each side of the
source at the same time. Another one was to perform the
measurements in different distances (so was the time) for the two
sides. The strangeness was tagged via strong interaction with
absorbers away from the creation point. The measured Strangeness
asymmetry is
\begin{eqnarray}
A(t_l, t_r) = \frac{I_{unlike}(t_l, t_r) - I_{like}(t_l,
t_r)}{I_{unlike}(t_l, t_r) + I_{like}(t_l, t_r)}\; .
\end{eqnarray}
Here, $I_{(un)like}$ means the (un)like strangeness event, which
are defined as
\begin{eqnarray}
I_{like}(t_l, t_r) & = & P(K^0,t_l;K^0,t_r)+
P(\bar{K}^0,t_l;\bar{K}^0,t_r) \; , \\ I_{unlike}(t_l, t_r) & = &
P(K^0,t_l;\bar{K}^0,t_r)+ P(\bar{K}^0,t_l;K^0,t_r)\; .
\end{eqnarray}
From Fig.\ref{cplear} one notices that the non-separability
hypothesis of QM is strongly favoured by experiment.

The $B^0\bar{B^0}$ entangled system produced at the $\Upsilon(4S)$
resonance has also been measured in the B-factory \cite{b0}. The
wave function of $\Upsilon(4S) \rightarrow B^0\bar{B^0}$ has the
same formalism as the spin singlet:
\begin{eqnarray}
|\Upsilon\rangle = \frac{1}{\sqrt{2}}\{ |B^0\rangle_l
|\bar{B^0}\rangle_r - |\bar{B^0}\rangle_l |B^0\rangle_r \}\; .
\end{eqnarray}
\begin{figure}
\begin{center}
\includegraphics[width=8cm,height=6cm]{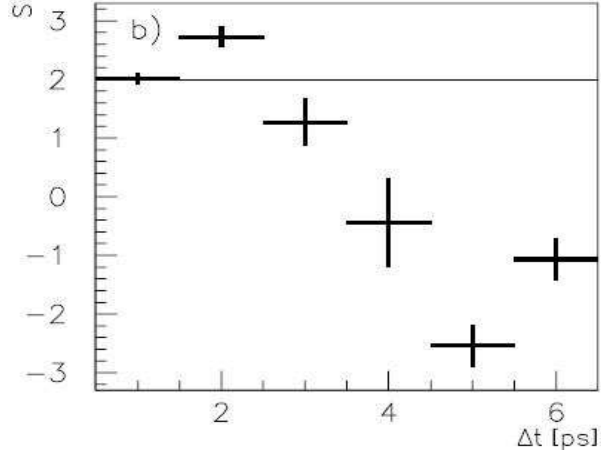}
\end{center}
\caption{\label{bb} \small The experiment result of the violation of
inequality \cite{b0}. The horizontal axis refers to $\Delta t$ and
the vertical axis to the S. The LHVTs limit of 2 is shown by the
solid line.}
\end{figure}
Here, the quantum number of flavor plays the role of spin
polarization in the spin correlation system. In the first method,
i.e. with fixed quasi-spin and free in time, the correlation
function for $B^0\bar{B^0}$ system reads
\begin{eqnarray}
E(t_l, t_r) = -e^{-\frac{2t' + \Delta t}{\tau_B}} \cos(\Delta m_d
\Delta t)\; ,
\end{eqnarray}
where $\Delta m_d$ characterizes the $B^0-\bar{B^0}$ mixing,
$\tau_B$ is the $B^0$ decay mean time, $t'=\min(t_l,t_r)$, and
$\Delta t = |t_l - t_r|$. Normalizing the above correlation function
by the undecayed $B^0$ pairs, one then gets the correlation function
as
\begin{eqnarray}
E_R(\Delta t) = - \cos( \Delta m_d \Delta t)\; .
\label{reducecorrelation}
\end{eqnarray}
Put it into the Bell-CHSH inequality, one can get the violation
parameter \cite{b0}
\begin{eqnarray}
S(\Delta t) = 3E_R(\Delta t) - E_R(3\Delta t) \leq 2\; .
\end{eqnarray}
The experiment, which based on the data sample of $80\times10^6$
$\Upsilon(4S) \rightarrow B\bar{B}$ decays at Belle detector at
the KEKB asymmetric collider in Japan, tells $S=2.725\pm
0.167_{stat}$. It is obviously a result violating the Bell
inequality, as shown in Fig.\ref{bb}.

In spite of the achievements in the high energy physics experiments
mentioned in above, theoretically, debate on whether the quasi-spin
of unstable particle can give a genuine test of LHVTs or not remains
\cite{verified}. If the neutral mesons are stable, the analogy of
quasi-spin with spin would be perfect. However, in reality the
unstable particles may decay, and hence, in principle on should
include the Hilbert space of all decay products as well
\cite{Bertlmann}. By the unitary time evolution of the unstable
state, some information may lose into the decay products. In
addition, there is another major difference between real- and
quasi-spin systems. For the former, one can detect arbitrarily the
spin state $\alpha |+\rangle + \beta |-\rangle$; however, it is not
true in the quasi-spin, the mesons pairs, system. This difference
may induce problem for the neutral meson systems. That is, the
passive measurement nature of quasi-spin meson system makes the
possibility to choose freely the quasi-spin among alternative setups
lost. In CPLEAR experiment the active measurement requirement is
fulfilled, because the neutral kaon meson is identified through
strong interaction with the absorber. While in the B meson case,
there is no way for experimenter to force B-meson to decay at a
given instant $t_l$ or $t_{l'}$ \cite{verified}. As of the unitary
condition, the Eq.(\ref{reducecorrelation}) for B-meson system is
different from Eq.(\ref{kaoncorrelation}). For B-meson system, it is
normalized by the undecayed $B^0\bar{B^0}$ pairs, which, like in the
photon case, asks some additional assumptions to make the
correlation to be comparable to what required by the Bell inequality
because of the detecting efficiency.

Recently, the $B_S^0 \bar{B}^0_S$ pair production is observed in
experiment in  $\Upsilon(5S)$ decays \cite{BS1,BS2}. $B^0_S$ meson
has a suitable $x$ value for the violation of CHSH inequality, even
if the interplay of weak interaction is considered, we expect that
measurement on $B_S^0 \bar{B}^0_S$ mixing in the future may give
another notable test of the QM correlation.

\subsection{Some Novel Ideas in Testing LHVT in High Energy Physics}

Recently, based on the Hardy's approach Bramon and Garbarino propose
a new scheme \cite{prl88,prl89} to test the local realism by virtue
of entangled neutral kaons. After neglecting the small CP-violation
effect, the initial $K_S K_L$ pair from $\phi$ decay, or
proton-antiproton annihilation, is the same as (\ref{kaon-no-cp}),
i.e.
\begin{eqnarray}
|\phi(T=0)\rangle=\frac{1}{\sqrt{2}}[K_SK_L-K_LK_S]\ ,\label{phi0}
\end{eqnarray}
where $K_S=(K^0+\bar{K^0})/\sqrt{2}$ and
$K_L=(K^0-\bar{K^0})/\sqrt{2}$ are the mass eigenstates of the $K$
mesons. One of the key points in using kaon system to test the
LHVTs is to generate a nonmaximally entangled (asymmetric) state.
That is
\begin{eqnarray}
|\phi(T)\rangle=\frac{1}{\sqrt{2+|R|^{\,2}}}\,
[K_SK_L-K_LK_S-re^{-i(m_L-m_S)\,T+[(\Gamma_S-\Gamma_L)/2]
T}K_LK_L]\ .\label{jpsiT}
\end{eqnarray}
Here, $r$ is the regeneration parameter to be the order of
magnitude $10^{-3}$ \cite{prl89}; $\Gamma_L$ and $\Gamma_S$ are
the $K_L$ and $K_S$ decay widths, respectively; T is the evolution
time of kaons after their production. Technically, this asymmetric
state can be achieved by placing a thin regenerator close to the
$\phi$ decay point \cite{prl88}.

Four specific transition probabilities for joint measurements from
QM take the following forms:
\begin{eqnarray}
P_{QM}(K^0,\bar{K^0}) & \equiv & |\langle
K^0\bar{K^0}|\phi(T) \rangle|^{\ 2}\nonumber \\
 & = & \frac{|2+\mathrm{R}e^{i\varphi}|^{\,2}}
 {4(2+|R|^2)}\ ,
 \label{tp1}
\end{eqnarray}
\begin{eqnarray}
P_{QM}(K^0,K_L) & \equiv & |\langle K^0 K_L|\phi
(T)\rangle|^{\ 2}\nonumber \\
& = & \frac{|1 + \mathrm{R}e^{i\varphi}|^{\,2}}{2(2 + |R|^2)}\ ,
 \label{tp2}
\end{eqnarray}
\begin{eqnarray}
P_{QM}(K_L,\bar{K^0}) & \equiv & |
\langle K_L\bar{K_0}|\phi(T)\rangle|^{\ 2}\nonumber \\
& = & \frac{|1+\mathrm{R}e^{i\varphi}|^{\,2}}{2(2+|R|^2)}\ ,
 \label{tp3}
\end{eqnarray}
\begin{eqnarray}
P_{QM}(K_SK_S)& \equiv & |\langle K_SK_S|\phi(T)\rangle|^{\ 2}= 0
\ ,
 \label{tp4}
\end{eqnarray}
where $\mathrm{R}=-|R|=-|r|e^{[(\Gamma_S-\Gamma_L)/2]T}$ and
$\varphi$ is the phase of $R$. In Ref. \cite{prl89} the special
case of $R=-1$ was considered, in which
\begin{eqnarray}
P_{QM}(K^0,\bar{K^0}) & = & 1/12\; , \label{4.1}\\ P_{QM}(K^0,K_L) &
= & 0\; , \label{4.2}\\ P_{QM}(K_L,\bar{K^0}) & = & 0\; , \label{4.3} \\
P_{QM}(K_S,K_S) & = & 0\; . \label{4.4}
\end{eqnarray}
From (\ref{jordan}) and in light of the arguments in Ref.
\cite{prl89}, in the following we demonstrate how LHVTs conflict
with QM in this case.

Suppose in a typical experiment, the strangeness on both sides at a
proper time T is measured. For example, a detection of $K^0$ on the
left side and $\bar{K^0}$ on the right side is achieved. We know
this may happen from Eq.(\ref{4.1}), and then we can infer from
(\ref{4.2}) that if the decay on the right hand side is observed,
the $K_S$ exits there for certain. In this case, according to
Einstein's argument the $K_S$ on the right side corresponds to a
physical reality. Similarly, if we have measured the lifetime of
kaon on the left side, according to (\ref{4.3}) one can confirm that
it should be $K_S$. In all, the non-zero probability of
$P_{QM}(K^0,\bar{K^0})$ leads to the non-zero probability of $K_S$
on both sides. However, due to EPR's criterion of ``physical
reality" this is in contradiction with Eq.(\ref{4.4}). This kind of
contradiction need a null measurement of the transition probability
of Eq.(\ref{4.4}) that cannot be strictly performed. Starting from
(\ref{eber}) Bramon \textit{et al.} derived out the following
Eberhard's inequality (EI) \cite{quant}, i.e.,
\begin{eqnarray}
H_{LR} \equiv \frac{P_{LR}(K^0,\bar{K^0})}{ P_{LR}(K^0,K_L) +
P_{LR}(K_S,K_S) + P_{LR}(K_L,\bar{K^0}) + P(K^0,U_{Lif}) +
P(U_{Lif},\bar{K^0}) } \leq 1 ,
\end{eqnarray}
where $P_{\rm {LR}}$ denotes the transition probability in LHVTs
with the subscripts LR symbolizing the local realism. $H_{LR}$ means
the local realistic value of the fraction which must less than 1
according to LHVTs. $U_{Lif}$ denotes the failures in lifetime
detection. In Ref.\cite{quant} the above inequality is used in
deducing the possible violation, which depends upon the restriction
of experimental efficiencies. Unlike the null measurement this
inequality can tolerate with the unsatisfied experimental
efficiencies.

For simplicity we consider an ideal case, in which the detection
efficiency of the kaon decays is 100 percent. Then the EI for the
kaon system takes the similar form as Eq.(\ref{eber})
\cite{Eberhard,garuccio}. It reads
\begin{eqnarray}
P_{LR}(K^0,\bar{K^0}) & \leq &
P_{LR}(K^0,K_L)+P_{LR}(K_S,K_S)+P_{LR}(K_L,\bar{K^0})\ .\label{ei}
\end{eqnarray}
\begin{figure}[t,m,u]
\begin{center}
\includegraphics[width=9cm,height=8cm]{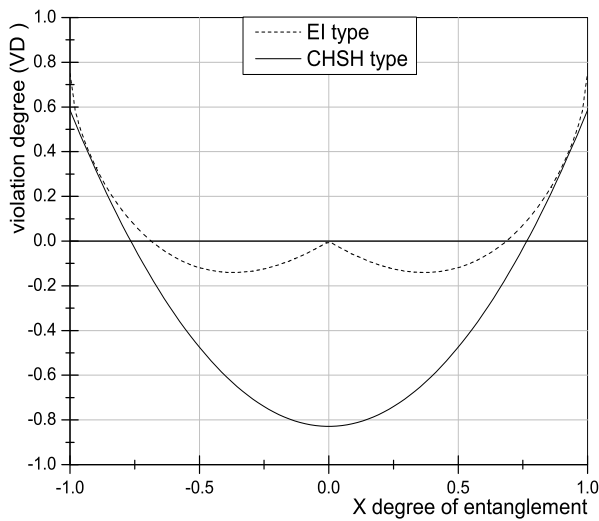}
\end{center}
\caption{\label{breaking1}\small The violation degree of the Bell
inequalities (the dashed line for EI type and the solid line for
CHSH type) in terms of the entanglement. Here, for the sake of
transparency, we make a coordinator exchange, the $C = 1-x^2$. The
magnitudes of $V\!\!D$ less than zero means the broken of the BIs. }
\end{figure}
For the case of QM, substituting equations (\ref{tp1}) -
(\ref{tp4}) into the inequality (\ref{ei}) and assuming $\varphi =
0$, we have
\begin{eqnarray}
\frac{(2+\mathrm{R})^2}{4(2+\mathrm{R}^2)} & \leq &
\frac{(1+\mathrm{R})^2}{2(2+\mathrm{R}^2)}\; +\; 0\; + \;
\frac{(1+\mathrm{R})^2}{2(2+\mathrm{R}^2)}\ . \label{ei1}
\end{eqnarray}
The above inequality is apparently violated by QM while $\mathrm{R}
= -1$. In Ref. \cite{ourself} we generalized the method used in
\cite{prl89} to heavy quarkonium. This straightforward
generalization however leads to some new observations of the
nonlocal property. Upon further analyzing the R value when it gives
violation of Eq.(\ref{ei1}), we find out that there exist a period
of time during which the violation became larger through time
evolution. In quantum information theory the entanglement property
of two qubits pure state are well understood, which can be
characterized by the concurrence $C$ \cite{concurrence}. We can also
see how the degree of entanglement evolves with time. Here,
according to the definition of concurrence we have
\begin{eqnarray}
C(J/\psi)=|\langle
J/\psi|\widetilde{J/\psi}\rangle|=\frac{2}{2+|R|^{\,2}}=
\frac{2}{2+|r|^{\,2}e^{(\Gamma_S-\Gamma_L)\,T}}\ , \label{ev1}
\end{eqnarray}
where $|\widetilde{J/\psi}\rangle = \sigma^1_y \sigma^2_y
|(J/\psi)^{*}\rangle\ $ and $\sigma^{1,\ 2}$ are Pauli matrices. $C$
changes between null to unit for no entanglement and full
entanglement. Eq.(\ref{ev1}) shows that the state become less
entangled with the time evolution. So, considering of (\ref{ei1}) we
realize that the violation of it does not decrease monotonously with
the degree of entanglement. To clarify this phenomenon we express
the violation degree ($V\!\!D$) of the inequalities (left side minus
the right side) in term of $C$ and compare it with the usual CHSH
inequality \cite{CHSH}. In Figure \ref{breaking1} different $V\!\!D$
behaviors of CHSH's and Eberhard's inequalities are presented. For
CHSH case, the $V\!\!D_{\rm CHSH}$ is obtained in the same condition
as the maximal violation happens in the full entanglement, the $C =
1$. We have:
\begin{eqnarray}
V\!\!D_{CHSH} = \sqrt{2}(1+C)-2\ .
\end{eqnarray}
In fact, the above $V\!\!D_{\rm CHSH}$ can be deduced from the
results given in Refs. \cite{gisin,kar,degree of two}. For EI
case,
\begin{eqnarray}
V\!\!D_{EI}=\frac{-3(1-C)+2\sqrt{2}\sqrt{C-C^2}}{4}\ .
\end{eqnarray}
Here,  in EI the counterintuitive quantum effect shows up, i.e. the
less entanglement corresponding to a larger $V\!\!D$ in some region
(see Fig.\ref{breaking1}). It is worthy to notice that with the time
evolution, when $R$ becomes less than $-\frac{4}{3}$, the QM and
LHVTs both satisfy the inequality (\ref{ei}). Thus give a certain
asymmetrically entangled state, the Hardy state \cite{hardy}, the QM
and LHVTs can be well distinguished from the EI in the region of
$R\in[-4/3,0)$.

In a recent work \cite{brachingratio}, an improved measurement of
branching ratio $B(J/\psi \rightarrow K_S^0K_L^0) = (1.80 \pm 0.04
\pm 0.13)\times 10^{-4}$ is reported, which is significant larger
than previous ones. Entangled kaon pairs from heavy quarkonium
decays can be easily space-likely separated. Thus, little
evolution time T will guarantee the locality condition
\cite{ourself}, and hence enables us to test the full range of $R$
and so the peculiar quantum effects. It is promising and
worthwhile to implement such test in future tau-charm factories,
because of both the experimental feasibility and theoretical
importance.

\section{Conclusions}

In this article we present a brief review of EPR paradox related
studies in high energy physics. To make it self-contained, we also
present some basic materials on the history of EPR paradox and
experimental realizations, for instance in optics, though our main
concern in this work is on the test of LHVTs in high energy physics
experiment. The questions and hopes are presented and discussed. The
study of BI and quantum correlation in high energy physics in fact
has experienced a long time, and in this article it is impossible
for us to cover every aspect of developments in this subject. For
instance, in the kaon system there exist some different approaches
in the study \cite{caban-Walczak-r}. On this respect, readers may
refer to Refs.\cite{caban-Walczak-l, Bertlmann-Hiesmayr} and
references therein. Noticing that there must be some important
researches which are neglected and not referred in this work, we
fell sorry for those authors.

The developments in the study of Bell inequalities and quantum
information theory are very important for people to further
understand the elusive nature of quantum phenomena. Investigation on
testing the validity of LHVT in high energy physics is still an
active and intriguing topic. The study in turn also stimulates some
new experimental methods in high energy physics. Because in high
energy physics the elementary particles are just the quanta, which
obey the quantum theory, to test the theory in this regime looks
unique. To this aim, one can imagine there is still large capacity
for high energy physics to play a more important role in the future.
\\

\section*{ Acknowledgments}

The work was supported in part by the Natural Science Foundation of
China and by the Scientific Research Fund of GUCAS (NO. 055101BM03).

\end{document}